\documentclass[dvips,aoas,preprint]{imsart}
\RequirePackage[OT1]{fontenc}
\RequirePackage{amsthm,amsmath}

\newtheorem{corollary}{Corollary}
\RequirePackage{multirow}
\RequirePackage{ctable}
\RequirePackage{apacite}
\RequirePackage{imsart}
\usepackage{graphicx}
\usepackage{amssymb}
\usepackage{booktabs}
\usepackage{subfigure}

\startlocaldefs
\numberwithin{equation}{section}
\theoremstyle{plain}

\endlocaldefs

\begin{document}

\begin{frontmatter}

\title{On Dealing with Censored Largest Observations under Weighted Least Squares}
\runtitle{Dealing with Censored Largest Observations under WLS}


\author{\fnms{Md Hasinur Rahaman} \snm{Khan}\corref{t1}\ead[label=e1]{hasinur@isrt.ac.bd}}
\address{Applied Statistics\\Institute of Statistical Research and Training\\University of Dhaka\\Dhaka 1000\\\printead{e1}}
\affiliation{University of Dhaka}
\and
\author{\fnms{J. Ewart H.} \snm{Shaw}\ead[label=e2]{Ewart.Shaw@warwick.ac.uk}}
\address{Department of Statistics\\University of Warwick\\Coventry CV4 7AL\\\printead{e2}}
\affiliation{University of Warwick}


\runauthor{Khan, MHR and Shaw, JEH}

\begin{abstract}
When observations are subject to right censoring,
weighted least squares with appropriate weights (to adjust for censoring)
is sometimes used for parameter estimation.
With Stute's weighted least squares method,
when the 
largest observation is censored ($Y_{(n)}^+$),
it is natural to apply the redistribution to the right algorithm
of Efron (1967). However, Efron's redistribution algorithm can lead to bias and inefficiency in estimation.
This study explains the issues clearly and proposes some alternative ways of treating $Y_{(n)}^+$. The first four proposed approaches are based on the well known Buckley--James (1979) method of imputation with the Efron's tail correction and the last approach is indirectly based on a general mean
imputation technique in literature. All the new schemes use penalized weighted least squares optimized by quadratic programming implemented with the accelerated failure time models. Furthermore, two novel additional imputation approaches are proposed to impute the tail tied censored observations that are often found in survival analysis with heavy censoring. Several simulation studies and real data analysis demonstrated that the proposed approaches generally outperform Efron's redistribution approach and lead to considerably smaller mean squared error and bias estimates.
\end{abstract}


\begin{keyword}
\kwd{Accelerated failure time (AFT) model}
\kwd{Efron's tail correction}
\kwd{Right censoring}
\kwd{Stute's weighted least squares}
\end{keyword}

\end{frontmatter}

\section{Introduction}
The accelerated failure time (AFT) model is a linear regression model
where the response variable is
usually the logarithm of the failure time [Kalbfleisch and Prentice (2002\nocite{kalb:pren:02:book})].
Let $Y_{(1)}, \cdots,Y_{(n)}$ be the ordered logarithm of survival times,
and $\delta_{(1)},\cdots,\delta_{(n)}$ are the corresponding censoring indicators.
Then the AFT model is defined by
\begin{equation}
Y_i=\alpha+X_i^T\beta +\varepsilon_i,~~i=1,\cdots,n,
\label{eq:aftmW}
\end{equation}
where $Y_i=\log\,(T_i)$,
$X_i$
is the covariate vector,
$\alpha$ is the intercept term, $\beta$ is the unknown $p\times 1$
vector of true regression coefficients and the $\varepsilon_i$'s are
independent and identically distributed random variables whose
common distribution may take a parametric form,
or may be unspecified, with zero mean and bounded variance.
For example, a log-normal AFT model is obtained if the error term $\varepsilon$ is normally distributed. As a result, we have a log linear type model that appears to be similar to the standard linear model that is typically estimated using ordinary least squares (OLS). But this AFT model can not
be solved using OLS because it can not handle censored data. The
trick to handle censored data turns out to introduce weighted least
squares method, where weights are used to account for censoring.

There are many studies where weighted least squares is used for AFT
models [Huang, Ma and Xie (2006\nocite{hua:ma:xie:06:regu}), Hu and Rao (2010), Khan and Shaw(2013\nocite{Kha:sha:13:Variable}\nocite{Kha:sha:13:VariableSelec})]. The AFT model (\ref{eq:aftmW}) is solved using a penalized version of Stute's weighted least squares method (SWLS) [Stute (1993, 1996\nocite{stut:93:consist}\nocite{stut:96:distrib})]. The SWLS estimate $\hat{\theta}=(\hat{\alpha},
\hat{\beta})$ of $\theta=(\alpha, \beta)$ is defined by
\begin{equation}
\hat{\theta}=\begin{array}{c}\arg\min\ \\ \theta
\end{array}\left[\frac{1}{2} \sum_{i=1}^{n}w_{i}\,(Y_{(i)}-\alpha-
X_{(i)}^T\beta)^2\right], \label{eq:wlseW}
\end{equation}
where $\frac{1}{2}$ is a normalizing constraint for
convenience, and the $w_{i}$'s are the weights which are typically
determined by two methods in the literature. One is called inverse
probability of censoring weighting (IPCW) and the other is called
Kaplan$-$Meier weight which is based on the jumps of the K$-$M
estimator. The IPCW approach is used in many studies in survival
analysis [e.g.~Robins and Finkelstein (2000\nocite{robi:fink:00:correct}), Satten and Datta (2001\nocite{satt:Datt:01:theKap})]. The K$-$M weighting approach is
also widely used in many studies such as
Stute (1993, 1994, 1996\nocite{stut:93:consist}\nocite{stut:96:thejack_var}\nocite{stut:96:distrib}),
Stute and Wang (1994\nocite{stut:94b:thejack}),
Hu and Rao (2010\nocite{hu:rao:2010:sparse}),
Khan and Shaw (2013\nocite{Kha:sha:13:Variable}).
The SWLS method in Equation (\ref{eq:wlseW}) uses the K$-$M weights to account for censoring.

The data consist of $(T_i^{*},\delta_i, X_i)$, $(i=1,\cdots,n)$, where $t^*_i=\min\,(t_i,c_i)$ and $\delta_i=I(t_i\leq c_i)$
where $t_i$ and $c_i$ represent the realization of the random
variables $T_i$ and $C_i$ respectively. Let $d_j$ be the number
of individuals who fail at time $t_j$ and $e_j$ be the number of
individuals censored at time $t_j$.
Then the K$-$M estimate of
$S(t)=\mbox{P}(T_i>t)$ is defined as
\begin{equation}
\widehat{S}(t)=\prod_{\{j:t_j\leq t\}}\left(1-\frac{d_j}{r_j}\right),
\label{eq:KMW1}
\end{equation}
where $r_j=\sum_{i=1}^{n}I(t_j\geq t)$ is the number of individuals at risk at time $t$.
In Stute (1993, 1996)\nocite{stut:93:consist}
\nocite{stut:96:distrib} the K$-$M weights are defined as follows
\begin{equation}
w_{1}=\frac{\delta_{(1)}}{n},~
w_{i}=\frac{\delta_{(i)}}{n-i+1}\prod_{j=1}^{i-1}\Big(\frac{n-j}{n-j+1}\Big)^{\delta_{(j)}},~~j=2,\cdots,n.
\label{eq:kmweights2}
\end{equation}
Note that this assigns $\frac{1}{n}$ weight to each
observation if all observations in the dataset are uncensored.

As can be observed, the K--M weighting method (\ref{eq:kmweights2}) gives zero weight to the censored
observations $Y_{(.)}^+$. The method also gives zero weight to the
largest observation if it is censored $\delta_{(n)}=0$.
Furthermore we know from the definition of the K$-$M
estimator (\ref{eq:KMW1}) that the K$-$M
estimator $\widehat{S}(t)$ is not defined for $Y> Y_{(n)}^+$ i.e.
\begin{equation}
\widehat{S}(t)=\begin{cases}\prod_{\{j|t_j\leq t\}}\left(1-\frac{d_j}{r_j}\right),&\mbox{for}~~~ t\leq T_{(n)}\\
\begin{cases}0,\qquad\qquad~\mbox{if}~\delta_{(n)}=1\\\mbox{undefined},~~\mbox{if}~\delta_{(n)}=0\end{cases},&\mbox{for}~~~ t> T_{(n)}.\end{cases}
\label{eq:KMW}
\end{equation}

This problem is usually solved by making a tail
correction to the K$-$M estimator. The correction is known as the redistribution
to the right algorithm proposed by Efron (1967\nocite{efro:67:the}).
Under this approach $\delta_{(n)}=0$ is reclassified as $\delta_{(n)}=1$ so that
the K$-$M estimator drops to zero at $Y_{(n)}^+$ and beyond, leading to obtaining proper (weights adding to one) weighting scheme.
Several published studies give zero weight
to the observation $Y_{(n)}^+$
[e.g.~Huang, Ma and Xie (2006\nocite{hua:ma:xie:06:regu}), Datta, Le-Rademacher and Datta (2007\nocite{dat:le:dat:07:pred})].
This has adverse consequences, as shown below.

\subsection{An Illustration}
In this study we consider only the K$-$M weights.
\begin{table}[h]
\centering
\caption{Survival times for 10 rats and their corresponding K--M weights with tail correction ($w_1$) and without tail correction ($w_0$)}
\scalebox{1}{\begin{tabular}{@{}lcccccccccc@{}}\toprule\toprule
Rat&1&2&3&4&5&6&7&8&9&10\\\midrule[.2ex]
$T_i$&9&13&13+&18&23&28+&31&34+&45&48+\\
$w_0$&0.100&0.100&0.000&0.114&0.114&0.000&0.143&0.000&0.214&0.000\\
$w_1$&0.100&0.100&0.000&0.114&0.114&0.000&0.143&0.000&0.214&0.214\\[1ex]
\bottomrule\bottomrule
\end{tabular}}
\label{tab:weightsex1}
\end{table}
Table \ref{tab:weightsex1} presents the hypothetical survival times
for ten rats, subject to right censoring. The table also presents the weight calculations with ($w_1$) and without ($w_0$) tail correction.
%

The Table \ref{tab:weightsex1} reveals that weighting without tail correction causes improper
weighting scheme. For the AFT model analyzed by weighted least
squares as defined by Equation (\ref{eq:wlseW}),
the improper weights will not contribute to the term $\frac{1}{2}
\sum_{i=1}^{n}w_{i}\,(Y_{(i)}-\alpha- X_{(i)}^T\beta)^2$ for the
observation $Y_{(n)}^+$.
Since the term $w_{i}\,(Y_{(i)}-\alpha- X_{(i)}^T\beta)^2$ is
non-negative this leads to a smaller value of weighted residual
squares compared to its actual value, resulting in a biased
estimate for $\beta$.
As the censoring percentage $P_{\%}$ increases,
the chance of getting the censored observation $Y^+_{(n)}$
also increases.


Therefore, both approaches with and without the tail correction
affect the underlying parameter estimation process,
giving biased and inefficient estimates in practice.
In the following section we introduce some
alternative options of dealing with the a censored largest
observation.
In the study we only consider datasets
where the largest observation is censored.

\section{Penalized SWLS}
Here we introduce a $\ell_2$ penalized WLS method to solve the AFT model (\ref{eq:aftmW}).
We first adjust $\mathbf{X}_{(i)}$ and $Y_{(i)}$ by centering them by their weighted means
\begin{equation*}
\bar{X}_{w}=\frac{\sum_{i=1}^n w_{i}\,X_{(i)}}{\sum_{i=1}^n
w_{i}},\qquad \bar{Y}_{w}=\frac{\sum_{i=1}^n
w_{i}\,Y_{(i)}}{\sum_{i=1}^n w_{i}}.
\end{equation*}
The weighted covariates and responses become
$X_{(i)}^w=(w_{i})^{1/2}(X_{(i)}-\bar{X}_{w})$ and
$Y_{(i)}^w=(w_{i})^{1/2}(Y_{(i)}-\bar{Y}_{w})$ respectively, giving
the weighted data $(Y_{(i)}^w,\delta_{(i)},X_{(i)}^w)$.
By replacing the original data $(Y_{(i)},\delta_{(i)},X_{(i)})$ with the weighted data, the objective function of the SWLS (\ref{eq:wlseW}) becomes
\begin{equation*}
\ell(\beta)=\frac{1}{2} \sum_{i=1}^{n}(Y_{(i)}^w-{X_{(i)}^w}^T\beta)^2.
\end{equation*}
Then, the ridge penalized estimator, $\hat{\beta}$, is the solution that minimizes
\begin{equation}
\ell(\beta)=\frac{1}{2} \sum_{i=1}^{n}(Y_{(i)}^w-{X_{(i)}^w}^T\beta)^2+\lambda_2\,\sum_{j=1}^p\beta_j^2, \label{eq:wlseW2}
\end{equation}
where $\lambda_2$ is the ridge penalty parameter.
The reason for choosing $\ell_2$ penalized estimation
is to deal with any collinearity among the covariates.
We use $\lambda_2=0.01\sqrt{2\log{p}}$ for the
log-normal AFT model because the $\sigma\sqrt{2\log{p}}$ term is a
natural adjustment for the number of variables (p) for model with
Gaussian noise [Candes and Tao (2007\nocite{cand:tao:07:thedantzig})].

We further develop this penalized WLS (\ref{eq:wlseW2}) in the spirit of the study
by Hu and Rao (2010\nocite{hu:rao:2010:sparse}).
The objective function of the modified penalized WLS
is defined in matrix form as below
\begin{equation}
\ell(\beta)=\frac{1}{2}(Y_{u}^w-X_{u}^w\beta)^T(Y_{u}^w-X_{u}^w\beta)+\frac{1}{2}\lambda_2\beta^T\beta~~
\text{subject to}~Y_{\bar{u}}^w\leq Y^w, \label{eq:wlseW3}
\end{equation}
where $Y_{u}^w$ and $X_{u}^w$ are the response variables and the covariates respectively both corresponding to the uncensored data and $Y^w$ is the associated unobserved log-failure times for censored observations. For censored data they are denoted by $Y_{\bar{u}}^w$ and $X_{\bar{u}}^w$ respectively.
The censoring constraints $Y_{\bar{u}}^w\leq Y^w$
arise from the right censoring assumption.
The optimization of Equation (\ref{eq:wlseW3}) is then carried out using a standard quadratic programming (QP) that has the form
\begin{equation}
\begin{array}{c}\arg\min\ \\ b \end{array}\left[-d^Tb+\frac{1}{2}b^TDb\right]\qquad \text{subject to}~ Ab \geq b_{0}.
\label{eq:qp}
\end{equation}
%

\section{Proposed Approaches for Imputing $Y_{(n)}^+$}
Let $T_{\bar{u}}$ be the true log failure times corresponding to the unobserved 
so that $T_{\bar{u}i}> Y_{\bar{u}i}$ for the $i$-th
censored observation.
We propose five approaches for imputing the
largest observation $Y_{\bar{u}(n)}$.
The first four approaches are based on the well known Buckley$-$James
(1979\nocite{buc:jam:79:lin}) method of imputation for censored observations with Efron's tail correction.
The last approach is indirectly based on the mean
imputation technique as discussed in Datta
(2005\nocite{datt:05:estimati}).
So, under the first four approaches,
the lifetimes are assumed to be modeled using the associated covariates but under the last approach there is no such assumption.

\subsection{Adding the Conditional Mean or Median}\label{subs:conMMA}
The key idea of the Buckley--James method for censored data is to replace the
censored observations (i.e.~$Y_{\bar{u}i}$) by their conditional expectations
given the corresponding censoring times and the covariates,
i.e.~$E(T_{\bar{u}i}|T_{\bar{u}i}>Y_{\bar{u}i},X_i)$.
Let $\xi_i$ is
the error term associated with the data $(T^*_i,\delta_i, X_i )$ i.e.~$\xi_i=Y_{i}-X_i^T\beta$ where $Y_i=\log\,(T^*_i)$ such that
solving the equation $\sum_{i=1}^n
X_i^T(Y_{i}-X_i^T\beta)=\sum_{i=1}^n X_i^T \xi_i=0$ yields the least squares estimates for $\beta$. According to the
Buckley$-$James method the quantity
$E(T_{\bar{u}i}|T_{\bar{u}i}>Y_{\bar{u}i},X_i)$ for the $i$-th censored
observation is calculated as
\begin{equation}
E(T_{\bar{u}i}|T_{\bar{u}i}>Y_{\bar{u}i},X_i)=Y_{\bar{u}i}+E(\varepsilon_i|\varepsilon_i>\xi_i,X_i).
\label{eq:censimput}
\end{equation}
We do not impute the largest observation, $Y_{\bar{u}(n)}$ using Equation (\ref{eq:censimput}), rather we add the conditional mean of ($\varepsilon|\varepsilon>\xi,X$) i.e.~$E(\varepsilon|\varepsilon>\xi)=\tau_m$ (say) or the conditional median of ($\varepsilon|\varepsilon>\xi,X$) i.e.~$\mbox{Median}(\varepsilon|\varepsilon>\xi)=\tau_{md}$ (say) to
$Y_{\bar{u}(n)}$. Here the quantity ($\varepsilon|\varepsilon>\xi,X$) is equivalent to ($\varepsilon|\varepsilon>\xi$)
since $\varepsilon\perp X$ in linear regression (\ref{eq:aftmW}).
The quantity $Y_{\bar{u}(n)}+\tau_m$ or $Y_{\bar{u}(n)}+\tau_{md}$
is therefore a reasonable estimate of the true log failure time
$T_{\bar{u}(n)}$ for the largest observation.

The quantity $\tau_m$ can be calculated by
\begin{align}
\tau_m=E(\varepsilon|\varepsilon>\xi)=\int_{\xi}^{\infty}\varepsilon\,
                                        \frac{\mbox{d}F(\varepsilon)}{1-F(\xi)},
\label{eq:ximean}
\end{align}
where $F(\cdot)$ is the distribution function.
Buckley and James (1979\nocite{buc:jam:79:lin}) show
that the above $F(\cdot)$ can be replaced by its Kaplan--Meier
estimate $\hat{F}(\cdot)$.
Using this idea Equation (\ref{eq:ximean})
can now be written as
\begin{align}
\tau_m=\sum_{j:\xi_j>\xi_i}\xi_j\,\frac{\Delta\hat{F}(\xi_j)}{1-\hat{F}(\xi_i)},
\label{eq:ximeanfinal}
\end{align}
where $\hat{F}$ is the Kaplan$-$Meier estimator of $F$ based on [($\xi_i,\,\delta_i),\, i=1,\cdots,n]$ i.e.,
\begin{align}
\widehat{F}(\xi_i)=1- \prod_{j:\xi_j>\xi_i}\left(1-\frac{\delta_i}{\sum_{j=1}^{n}1_{\{\xi_j\geq \,\xi_i\}}} \right).
\label{eq:bj-km}
\end{align}

The conditional median $\tau_{md}$ can be calculated from the following expression.

\begin{align}
\int_{\xi_i}^{\tau_{md}}\frac{\mbox{d}F(\varepsilon_i)}{1-F(\xi_i)}&=0.5.
\label{eq:ximedian}
\end{align}
After replacing $F(\cdot)$ by its K$-$M estimator $\hat{F}(\cdot)$
Equation (\ref{eq:ximedian}) can be written as
\begin{align}
\sum_{j:\xi_j>\xi_i} 
\frac{\Delta\hat{F}(\xi_j)}{1-\hat{F}(\xi_i)}=0.5.
\label{eq:ximedianfinal}
\end{align}

\subsection{Adding the Resampling-based Conditional Mean and Median}\label{subs:ReCMMA}
The approaches are similar to adding the conditional
mean and median as discussed in Section (\ref{subs:conMMA})
except that $\tau_{m}$ and $\tau_{md}$ are calculated
using a modified version of an iterative solution to the
Buckley$-$James estimating method [{Jin, Lin and Ying}(2006{\nocite{jin:lin:ying:06:Onleast}})] rather than the original
Buckley$-$James (1979\nocite{buc:jam:79:lin}) method.
We have followed the
iterative Buckley$-$James estimating method [{Jin, Lin and Ying}(2006{\nocite{jin:lin:ying:06:Onleast}})] along with the associated imputation technique because it provides
a class of consistent and asymptotically normal estimators. We have modified this iterative procedure by introducing
a quadratic programming based weighted least square estimator
as the initial estimator. Under this scheme we replace the
unobserved $Y_{\bar{u}(n)}$ by $Y_{\bar{u}(n)}+\tau^{\ast}_m$ or
$Y_{\bar{u}(n)}+\tau^{\ast}_{md}$ where $\tau^{\ast}_{m}$ and
$\tau^{\ast}_{md}$ are the resampling based conditional mean and
median calculated by
\begin{align}
\tau^{\ast}_m=\sum_{j:\xi_j>\xi_i}\xi_j\,\frac{\Delta\hat{F}^{\ast}(\xi_j)}{1-\hat{F}^{\ast}(\xi_i)},
\label{eq:newximeanfinal}
\end{align}
and
\begin{align}
\sum_{j:\xi_j>\xi_i}\frac{\Delta\hat{F}^{\ast}(\xi_j)}{1-\hat{F}^{\ast}(\xi_i)}=0.5
\label{eq:newximedianfinal}
\end{align}
respectively.
Here $\widehat{F}^{\ast}(\xi_i)$ is calculated using Equation (\ref{eq:bj-km}) based on the modified iterative
Buckley$-$James estimating method.
The procedure is described below.

Buckley and James (1979\nocite{buc:jam:79:lin}) replaces the
$i$-th censored $Y_{\bar{u}i}$ by
$E(T_{\bar{u}i}|T_{\bar{u}i}>Y_{\bar{u}i},X_i)$, yielding
                    \begin{align*}
                    \widehat{Y}_i(\beta)=\delta_{i}\,Y_i+(1-\delta_{i})\left[\int_{\xi_i}^{\infty}\varepsilon_i\,
                                        \frac{\mbox{d}\hat{F}(\varepsilon_i)}{1-\hat{F}(\xi_i)}+X_i^T\beta\right],
                    \label{eq:estYi}
                    \end{align*}
where $\hat{F}$ is the K$-$M estimator of $F$ based on the
transformed data ($\xi_i, \delta_i$) and that is defined by Equation
(\ref{eq:bj-km}). The associated Buckley$-$James estimating function
$U(\beta, b)=\sum_{i=1}^{n}(X_i-\bar{X})\{\hat{Y}_i(b)-X_i^T\beta\}$
is then defined by $U(\beta,
b)=\sum_{i=1}^{n}(X_i-\bar{X})\{\hat{Y}_i(b)-\bar{Y}(b)-(X_i-\bar{X}^T\beta)\}$
for $\bar{Y}(b)=n^{-1}\sum_{i=1}^n\hat{Y}_i(b)$. The Buckley$-$James estimator $\hat{\beta}_{bj}$ is the root of $U(\beta,\,\beta)=0$.  This gives the
following solution:
                    \begin{equation}
                    \beta=\mbox{L}(b)=\left\{\sum_{i=1}^{n}(X_i-\bar{X})^{\bigotimes 2}\right\}^{-1}\left[\sum_{i=1}^{n}(X_i-\bar{X})\{\hat{Y}_i(b)-\bar{Y}(b)\}\right],
                    \label{eq:newBeta}
                    \end{equation}
where $a^{\bigotimes 2}$ means $aa^T$ for a vector. The expression
(\ref{eq:newBeta}) leads to following iterative algorithm.
                    \begin{equation}
                    \hat{\beta}_{(m)}=\mbox{L}(\hat{\beta}_{(m-1)}),
                    \qquad m \ge 1.
                    \label{eq:BetaEst}
                    \end{equation}

In Equation (\ref{eq:BetaEst}) we set the initial estimator $\hat{\beta}_{(0)}$
to be the penalized weighted least square estimator $\hat{\beta}_{qp}$ that is obtained by optimizing the objective function as specified by the Equation (\ref{eq:wlseW3}).
The initial estimator $\hat{\beta}_{qp}$ is a consistent and asymptotically normal estimator
such as the Gehan-type rank estimator [{Jin, Lin and Ying}(2006{\nocite{jin:lin:ying:06:Onleast}})].
Therefore using $\hat{\beta}_{qp}$ as the initial estimator will satisfy the following corollary that immediately follows from {Jin, Lin and Ying}(2006{\nocite{jin:lin:ying:06:Onleast}}).
\begin{corollary}
The penalized weighted least squares estimator $\hat{\beta}_{qp}$ leads to a consistent and asymptotically normal $\hat{\beta}_{(m)}$ for each fixed $m$. In addition, $\hat{\beta}_{(m)}$ is a linear combination of $\hat{\beta}_{qp}$ and the Buckley$-$James estimator $\hat{\beta}_{bj}$ in that
\begin{equation*}
\widehat{\beta}_{(m)}= (I-D^{-1}A)^{m}\hat{\beta}_{qp}+\{I-(I-D^{-1}A)^{m} \}\hat{\beta}_{bj}+o_{p}\,(n^{-\frac{1}{2}})
\end{equation*}
where I is the identity matrix,
$D:=\lim_{n\rightarrow\infty}\,n^{-1}\,\sum_{i=1}^{n}(X_i-\bar{X})^{\bigotimes 2}$
is the usual slope matrix of the least-squares estimating function for the uncensored data,
and $A$ is a slope matrix of the Buckley$-$James estimation function
as defined in {Jin, Lin and Ying}(2006{\nocite{jin:lin:ying:06:Onleast}}).
\end{corollary}

{Jin, Lin and Ying}(2006{\nocite{jin:lin:ying:06:Onleast}}) also developed a resampling procedure to approximate the distribution of $\hat{\beta}_{(m)}$.
Under this procedure $n$ iid positive random variables $Z_i (i=1,\cdots,n)$ with $E(Z_i)=\mbox{var}(Z_i)=1$ are generated. Then define
                    \begin{align}
                    \widehat{F}^{\ast}(\xi_i)=1- \prod_{j:\xi_j>\xi_i}\left(1-\frac{Z_i\delta_i}{\sum_{j=1}^{n}Z_i1_{\{\xi_j\geq\,\xi_i\}}} \right),
                    \label{eq:newbj-km}
                    \end{align}
and
                    \begin{align*}
                    \widehat{Y}^{\ast}_i(\beta)=\delta_{i}\,Y_i+(1-\delta_{i})\left[\int_{\xi_i}^{\infty}\varepsilon_i\,
                                        \frac{\mbox{d}\hat{F}^{\ast}(\varepsilon_i)}{1-\hat{F}^{\ast}(\xi_i)}+X_i^T\beta\right],
                    \label{eq:estYi}
                    \end{align*}
and
                    \begin{equation}
                    \mbox{L}^{\ast}(b)=\left\{\sum_{i=1}^{n}Z_i(X_i-\bar{X})^{\bigotimes 2}\right\}^{-1}\left[\sum_{i=1}^{n}Z_i(X_i-\bar{X})\{\hat{Y}^{\ast}_i(b)-\bar{Y}^{\ast}(b)\}\right].
                    \label{eq:newBetastar}
                    \end{equation}
Equation (\ref{eq:newBetastar}) then leads to an iterative process $\hat{\beta}^{\ast}_{(m)}=\mbox{L}^{\ast}(\hat{\beta}^{\ast}_{(m-1)}),\,\, m \ge 1$.
The initial value $\hat{\beta}^{\ast}_{(0)}$ of the iteration process becomes $\hat{\beta}^{\ast}_{qp}$ which is the optimized value of
\begin{equation}
\frac{1}{2}\,Z(Y_{u}^w-X_{u}^w\beta)^T(Y_{u}^w-X_{u}^w\beta)+\frac{1}{2}\lambda_2\beta^T\beta,\qquad
\text{subject to}~ZY_{\bar{u}}^w\leq ZX_{\bar{u}}^w\beta.
\label{eq:newswlse3}
\end{equation}
This objective function (\ref{eq:newswlse3}) is obtained from the function as specified in Equation (\ref{eq:wlseW3}). For a given sample $(Z_1,\cdots,Z_n)$, the iteration procedure $\widehat{\beta}^{\ast}_{(k)}=\mbox{L}^{\ast}(\widehat{\beta}^{\ast}_{(k-1)})$ yields a $\widehat{\beta}^{\ast}_{(k)}\,(1\leq k\leq m)$. The empirical distribution of $\widehat{\beta}^{\ast}_{(m)}$ is based on a large number of realizations that are computed by repeatedly generating the random sample $(Z_1,\cdots,Z_n)$ a reasonable number of times. Then the empirical distribution is used to approximate the distribution of $\widehat{\beta}_{(m)}$ [{Jin, Lin and Ying}(2006{\nocite{jin:lin:ying:06:Onleast}})].

\subsection{Adding the Predicted Difference Quantity}
Suppose $Y^{m}_{\bar{u}i}$ is the modified failure time (imputed value) obtained using mean imputation technique to the censored $Y_{\bar{u}i}$.
Under mean imputation technique the censored observation, $Y_{\bar{u}i}$ is simply replaced with
the conditional expectation of $T_i$ given $T_i > C_i$.
The formula for estimating $Y^{m}_{\bar{u}i}$ is given by
                    \begin{equation}
                    Y^{m}_{\bar{u}i}=\{\hat{S}(C_i)\}^{-1}\sum_{t_{(r)}>C_i}\log{(t_{(r)})}\,\bigtriangleup
                    \widehat{S}(t_{(r)}),
                    \label{eq:imputedy}
                    \end{equation}
where $\hat{S}$ is the K$-$M estimator of the survival function of $T$ as defined by
Equation (\ref{eq:KMW1}), $\bigtriangleup \widehat{S}(t_{(r)})$ is the jump size of $\hat{S}$ at time $t_{(r)}$.
This mean imputation approach is used in many other studies [e.g.~Datta (2005\nocite{datt:05:estimati}), Datta, Le-Rademacher and Datta (2007\nocite{dat:le:dat:07:pred}), Engler and Li (2009)\nocite{eng:li:09:surv} etc].

We note that using mean imputation technique the modified failure time $Y^{m}_{\bar{u}(n)}$ to the largest observation $Y_{\bar{u}(n)}$ can not be computed since the K$-$M estimator is undefined for $\delta_{(n)}=0$. In particular, the quantity $\bigtriangleup\widehat{S}(t_{(r)})$ in Equation (\ref{eq:imputedy}) can not be calculated for the $n$-th observation $\delta_{(n)}=0$. This issue is clearly stated in Equation (\ref{eq:KMW}). Here we present a different strategy to impute $Y_{\bar{u}(n)}$. We assume that the mean imputation technique is used for imputing all other censored observations except the last largest censored observation. Let us assume that $\nu$ be a non-negative quantity such that $Y^{m}_{\bar{u}(n)}-Y_{\bar{u}(n)}=\nu$.
One can now estimate $\nu$ using many possible ways. Here we choose a very simple way that uses the
imputed values obtained by the above mean imputation approach. We estimate $\nu$ as a predicted value
based on the differences between the imputed values and the censoring times for all censored
observations except the largest observations.

\begin{figure}[ht]
\centering
\includegraphics [scale=0.60] {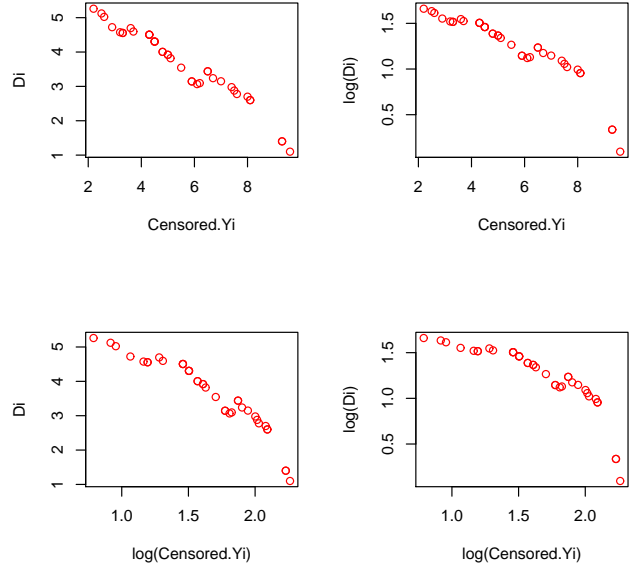}
\caption{Relationship between $D_i$ and $Y_{\bar{u}i}\,\, \mbox{for}\,\, i=1,\cdots,n_{\bar{u}i}-1$ based on the Larynx data} \label{fig:relationship}
\end{figure}

\begin{figure}[ht]
\includegraphics [scale=0.35] {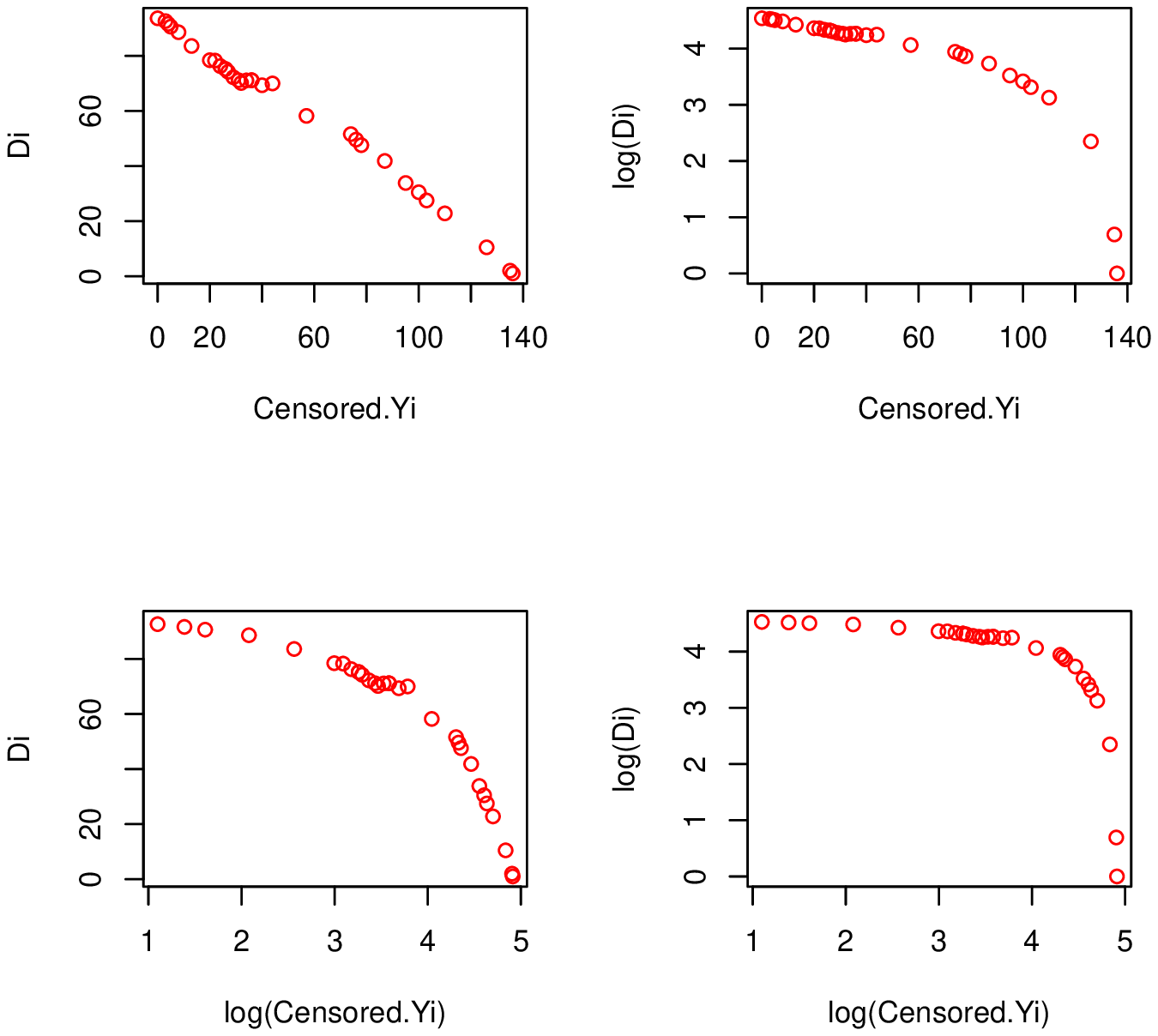}
\includegraphics [scale=0.35] {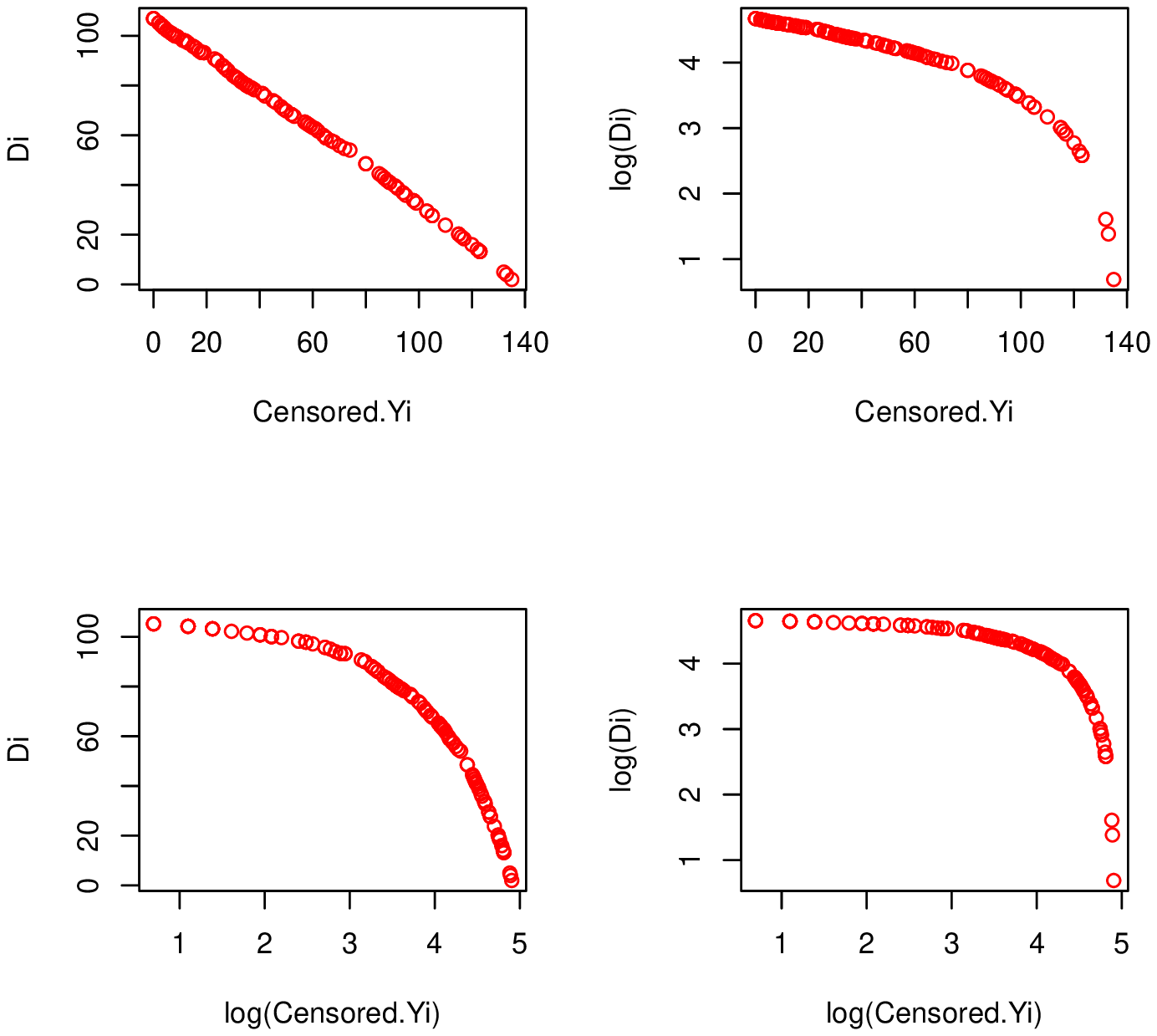}
\caption{Relationship between $D_i$ and $Y_{\bar{u}i}\,\, \mbox{for}\,\, i=1,\cdots,n_{\bar{u}i}-1$ is based on Channing House data. Left two column plots are based on male data and the right two column plots are based on female data}
\label{fig:ChanningHouseData}
\end{figure}

Suppose $D_i\,\, \mbox{for}\,\,i=1,\cdots,(n_{\bar{u}i}-1)$ represents the difference between the imputed value and the unobserved value for the $i$-th censored observation.
So, the quantity $\nu$ can be treated as a possible component of the $D$ family.
We examine the relationship between $D_i$ and $Y_{\bar{u}i}$ by conducting various numerical studies.
Figures \ref{fig:relationship} and \ref{fig:ChanningHouseData} show the most approximate
relationships between $D_i$ and $Y_{\bar{u}i}$ from two real datasets. Figure \ref{fig:relationship}
is based on the Larynx dataset [Kardaun (1983\nocite{Kard:83:Statis})]. Details are given in the real data analysis section.
Figure \ref{fig:ChanningHouseData} is based on the Channing House dataset [Hyde (1980\nocite{Hyde:1980:book})]
that also discussed in the section of real data analysis.
Both male and female data have heavy censoring toward the right.

%

Both Figures \ref{fig:relationship} and \ref{fig:ChanningHouseData} clearly suggest
a negative linear relationship between $D_i$ and $Y_{\bar{u}i}$.
The trend based on other transformations (logarithmic of either $D_i$ or $Y_{\bar{u}i}$ or both) appears to be nonlinear. Hence we set up a linear regression for $D_i$ on $Y_{\bar{u}i}$ which is given by
\begin{equation}
D_i=\tilde{\alpha}+Y_{\bar{u}i}\,\tilde{\beta} +\tilde{\varepsilon}_i,~~i=1,\cdots,(n_{\bar{u}i}-1),
\label{eq:relationship}
\end{equation}
where $\tilde{\alpha}$ is the intercept term, $\tilde{\beta}$ is the coefficient for the unobserved censored time, and $\tilde{\varepsilon}_i$ is the error term. We fit the model (\ref{eq:relationship}) with a WLS method that gives the objective function.
\begin{equation}
\sum_{i=1}^{n_{\bar{u}i}-1} \tilde{w}_i\,(D_i-\tilde{\alpha}-Y_{\bar{u}i}\,\tilde{\beta})^2,
\label{eq:relationship.weight}
\end{equation}
where $\tilde{w}_i$ are the data dependent weights. The weight for the $i$-th observation in Equation (\ref{eq:relationship.weight}) is chosen by $\{Y_{\bar{u}(n)}-Y_{\bar{u}i}\}^{-1}$. We choose WLS method for fitting model (\ref{eq:relationship}) because it is observed from Figure \ref{fig:relationship} that the $D_{i}$ occurs more frequently for the lower and middle censoring times than that for the higher censoring time. For this reason, perhaps the WLS method should be used for all future datasets. Finally the quantity $\nu$ is obtained by
                    \begin{equation}
                    \hat{\nu}=\begin{cases}\hat{D}(Y_{\bar{u}(n)}),\,\,~~\mbox{if}~\hat{D}(Y_{\bar{u}(n)})>0\\0,\qquad\qquad\,\mbox{if}~\hat{D}(Y_{\bar{u}(n)})\leq 0.\end{cases}
                    \label{eq:amd}
                    \end{equation}

\subsection{Proposed Additional Approaches for Imputing Tail Ties}\label{subs:addTail}
In the present of heavy censoring the proposed imputed methods are able to impute all the largest observations that are tied with only a single value.
In this case one might be interested in imputing the observations with different lifetimes as if they would have been observed. In order to acknowledge this issue we propose two alternative approaches for imputing the heavy tailed censoring observations. The approaches do not require to take the underlying covariates into account. The practical implications of such imputation techniques might be found in many fields such as economics, industry, life sciences etc. One approach is based on the technique of the predicted difference quantity. The other approach is based on the trend of the survival probability in the K--M curve.

\subsubsection{Iterative Procedure}
Let us assume that there are $m$ tied largest observations which are denoted, without loss of generality, by $Y_{\bar{u}(n_k)}$ for $k=1,\cdots,m$.
The first technique is an $m$ iterative procedure where the $k$-th observation is imputed using the predicted difference method after assuming that the $k$-th observation is the unique largest censored observation in the dataset. The computational procedure is summarized briefly as below
\begin{enumerate}
\item Compute the modified failure time using Equation (\ref{eq:imputedy}).
\item Set $\delta_{(n_k)}=1$ for any $1\leq k \leq m$.
\item Compute $\nu$ using Equation (\ref{eq:amd}).
\item Add the quantity $\nu$ found in Step 3 to $Y_{\bar{u}(n_k)}$.
\item Repeat Step 2 to 4 for $m-1$ times for imputing $m-1$ observations. The $\nu$ in Step 3 under each imputation is based on all modified failure times including the imputed values found in Step 4.
\end{enumerate}

\subsubsection{Extrapolation Procedure}\label{sssection:extrapolation}
Under this approach we first follow the trend of the K--M survival probabilities $\hat{S}(t)$ versus the lifetimes for the subjects.
If the trend for original K--M plot is not linear then we may first apply a transformation of the survival probability (e.g.~$[\hat{S}(t)]^{\psi}$ for suitable $\psi$).
When linear trend is established we fit a linear regression of lifetimes on $[\hat{S}(t)]^{\psi}$.
Now the lifetimes against the expected survival probabilities can easily be obtained using the fitted model.

\section{Estimation Procedures}
The performance of the proposed imputation approaches along with Efron's
(1967) redistribution technique is investigated with the AFT model
evaluated by the quadratic program based Stute's weighted least
squares method as discussed in Section (\ref{subs:ReCMMA}).
For convenience, let $W_{0}$ represents the estimation process
when no imputation is done for $Y_{\bar{u}(n)}$,
i.e.\ only Efron's (1967) redistribution algorithm is applied.
Let $W_{\tau_{m}}$, $W_{\tau_{md}}$, $W_{\tau^{\ast}_{m}}$, $W_{\tau^{\ast}_{md}}$, and $W_{\nu}$ represent estimation where $Y_{\bar{u}(n)}$ is imputed by adding the conditional mean, the conditional median, the resampling based conditional mean, the resampling based conditional median and the predicted difference quantity respectively.

\subsection{$W_0$: Efron's Approach}
\begin{enumerate}
\item Set $\delta_{(n)}=1$.
\item Solve Equation (\ref{eq:wlseW3}) using the QP approach (\ref{eq:qp}) to estimate $\beta$.
\end{enumerate}

\subsection{$W_{\tau_{m}}$: Conditional Mean Approach}
\begin{enumerate}
\item Set $\delta_{(n)}=1$ and solve Equation (\ref{eq:wlseW3}) using the QP approach (\ref{eq:qp}) to estimate $\beta$.
\item Compute weighted least squares errors based on the estimated $\beta$ from Step 1 and then calculate
      the K-M estimator of the errors using Equation (\ref{eq:bj-km}) and $\tau_{m}$ using Equation (\ref{eq:ximeanfinal}).
\item Add the quantity $\tau_{m}$ found in Step 2 to $Y_{\bar{u}(n)}$.
\item Get a new and improved estimate of $\beta$ based on modified $Y_{\bar{u}(n)}$ found in Step 3
      by solving Equation (\ref{eq:wlseW3}) using the QP approach (\ref{eq:qp}).
\end{enumerate}

\subsection{$W_{\tau_{md}}$: Conditional Median Approach} The
process of $W_{\tau_{md}}$ is similar to $W_{\tau_{m}}$ except
that it uses $\tau_{md}$ instead of $\tau_{m}$ in Step 2 and Step 3.

\subsection{$W_{\tau^{\ast}_{m}}$: Resampling based Conditional
Mean Approach}
\begin{enumerate}
\item Set $\delta_{(n)}=1$ and solve Equation (\ref{eq:newBetastar}) to estimate $\beta$.
\item Compute weighted least squares errors based on the estimated $\beta$ from Step 1 and then calculate
      the K-M estimator of the errors using the Equation (\ref{eq:bj-km}) and $\tau^{\ast}_{m}$ using Equation (\ref{eq:newximeanfinal}).
\item Add the quantity $\tau^{\ast}_{m}$ found in Step 2 to $Y_{\bar{u}(n)}$.
\item Get a new and improved estimate of $\beta$ based on modified $Y_{\bar{u}(n)}$ found in Step 3
      by solving Equation (\ref{eq:wlseW3}) using the QP approach (\ref{eq:qp}).
\end{enumerate}

\subsection{$W_{\tau^{\ast}_{md}}$: Resampling based Conditional
Median Approach} The process of $W_{\tau^{\ast}_{md}}$ is similar
to $W_{\tau^{\ast}_{m}}$ except that it uses $\tau^{\ast}_{md}$ instead of
$W_{\tau^{\ast}_{m}}$ in Step 2 and Step 3.

\subsection{$W_{\nu}$: Predicted Difference Quantity Approach}
\begin{enumerate}
\item Set $\delta_{(n)}=1$ and compute the modified failure time using Equation (\ref{eq:imputedy}).
\item Compute $\nu$ using Equation (\ref{eq:amd}).
\item Add the quantity $\nu$ found in Step 2 to $Y_{\bar{u}(n)}$.
\item Get an estimate of $\beta$ based on modified $Y_{\bar{u}(n)}$ found in Step 3
      by solving Equation (\ref{eq:wlseW3}) using the QP approach (\ref{eq:qp}).
\end{enumerate}

\section{Simulation Studies}
Here we investigates the performance of the imputation approaches
using a couple of simulation examples.
The datasets are simulated from the following log-normal AFT model, where
the largest observation is set to be censored (i.e.~$\delta_{(n)}=0$):
\begin{equation}
Y_i=\alpha+X_i^T\beta +\sigma \varepsilon_i,\qquad \varepsilon_i\sim N(0,1)~\text{for}~i=1,\cdots,n.
\label{eq:exampleW}
\end{equation}
The pairwise correlation ($r_{ij}$) between the $i$-th and $j$-th components of X is set to be $0.5^{|i-j|}$. The censoring times are generated using $U(a, 2a)$, where $a$ is chosen such that pre-specified censoring rates $P_{\%}$ are approximated.

\subsection{First Example}
We choose $n=100$, $p=5$, and $\sigma=1$, and $r_{ij}$ = 0 and 0.5, three censoring rates 30\%, 50\%, and 70\%. We choose $\beta_j=j+1$ for $j=1,\cdots,p$.
The bias, variance, and mean squared error (MSE) for $\beta$
are estimated by averaging the results from 1,000 runs.

\begin{table}\centering
\caption{Summary statistics for first simulation example $r_{ij}$ = 0. Comparison between the imputation approaches $W_0$: Efron's redistribution, $W_{\tau_{m}}$: conditional mean, $W_{\tau_{md}}$: conditional median, $W_{\tau^{\ast}_{m}}$: resampling based conditional mean, $W_{\tau^{\ast}_{md}}$: resampling based conditional median, and $W_{\nu}$: predicted difference quantity.} \scalebox{0.50}{
\begin{tabular}{@{}lccccccccccccccccccccc@{}}\toprule\toprule
&\multicolumn{6}{c}{$P_{\%}=30$}&&\multicolumn{6}{c}{$P_{\%}=50$}&&\multicolumn{6}{c}{$P_{\%}=70$}\\
\cmidrule{2-7}\cmidrule{9-14}\cmidrule{16-21}&$W_0$&$W_{\tau_{m}}$&$W_{\tau_{md}}$&$W_{\tau^{\ast}_{m}}$&$W_{\tau^{\ast}_{md}}$&$W_{\nu}$
&&$W_0$&$W_{\tau_{m}}$&$W_{\tau_{md}}$&$W_{\tau^{\ast}_{m}}$&$W_{\tau^{\ast}_{md}}$&$W_{\nu}$&&$W_0$&$W_{\tau_{m}}$&$W_{\tau_{md}}$&$W_{\tau^{\ast}_{m}}$&$W_{\tau^{\ast}_{md}}$&$W_{\nu}$\\
\hline
Bias&&&&&&&&&\\
$\beta_1$& 0.391&0.363&0.380&0.338&0.367&0.408&&0.467&0.402&0.450&0.403&0.430&0.472&&0.729& 0.595&0.665&0.488&0.588&0.709\\
$\beta_2$& 0.655 &0.616&0.639&0.605&0.622&0.678&&0.629&0.535&0.596&0.498&0.562& 0.636&&0.958&0.733&0.841&0.678&0.716&0.942\\
$\beta_3$& 0.839 &0.789&0.820&0.777&0.799&0.867&&0.877&0.761&0.841&0.748&0.793& 0.885&&1.299&1.030&1.143&0.921&0.998&1.270\\
$\beta_4$& 0.988 &0.935&0.963&0.911&0.941&1.018&&0.981&0.835&0.940&0.806&0.886& 0.986&&1.539&1.176&1.361&1.059&1.196&1.494\\
$\beta_5$& 1.226 &1.160&1.201&1.142&1.177&1.258&&1.271&1.087&1.225&1.065&1.162& 1.282&&2.193&1.812&1.992&1.642&1.798&2.151\\
\hline
Variance&&&&&&&&&\\
$\beta_1$& 0.154 &0.156&0.155&0.162&0.159&0.152&&0.283&0.348&0.290&0.328&0.296&0.291&&0.483&0.632&0.523&0.890&0.594&0.520\\
$\beta_2$& 0.174 &0.178&0.178&0.187&0.181&0.174&&0.282&0.324&0.285&0.331&0.294&0.290&&0.631&0.742&0.645&0.828&0.664&0.657\\
$\beta_3$& 0.142&0.146&0.143&0.146&0.143&0.144&&0.279&0.319&0.273&0.327&0.268&0.291&&0.536&0.666&0.561&0.774&0.581&0.581\\
$\beta_4$& 0.182 &0.181&0.181&0.190&0.182&0.180&&0.276&0.316&0.268&0.371&0.273&0.301&&0.618&0.628&0.543&0.665&0.550&0.616\\
$\beta_5$& 0.160 &0.164&0.161&0.164&0.158&0.166&&0.246&0.281&0.239&0.274&0.236&0.277&&0.620&0.690&0.561&0.784&0.605&0.659\\
\hline
MSE&&&&&&&&&\\
$\beta_1$&0.305 &0.286&0.298&0.275&0.292&0.317&&0.499&0.506&0.489&0.487&0.478& 0.511&&1.009&0.980&0.959&1.119&0.934&1.017\\
$\beta_2$&0.601 &0.556&0.585&0.550&0.567&0.632&&0.675&0.607&0.638&0.576&0.607& 0.692&&1.543&1.273&1.347&1.280&1.171&1.539\\
$\beta_3$&0.844 &0.786&0.814&0.748&0.781&0.894&&1.046&0.895&0.978&0.883&0.895& 1.072&&2.219&1.721&1.861&1.614&1.571&2.189\\
$\beta_4$&1.157 &1.054&1.107&1.019&1.067&1.213&&1.237&1.010&1.148&1.016&1.055& 1.270&&2.981&2.004&2.389&1.779&1.975&2.842\\
$\beta_5$&1.663 &1.508&1.601&1.465&1.542&1.746&&1.860&1.460&1.737&1.406&1.584& 1.918&&5.422&3.969&4.525&3.472&3.832&5.277\\[1ex]
\bottomrule\bottomrule
\end{tabular}}
\label{tab:weightingEX1}
\end{table}


The results for uncorrelated and correlated datasets are reported in Table \ref{tab:weightingEX1} and \ref{tab:weightingEX2} respectively. The results generally suggest that the resampling based conditional mean adding $W_{\tau^{\ast}_{m}}$ and the resampling based conditional median adding $W_{\tau^{\ast}_{md}}$ provide the smallest MSE in particular, smaller than Efron's approach $W_0$ at all censoring levels. They seem to provide generally lower bias except for $\beta_1$ and $\beta_2$ in correlated case.

We also find that at lower and medium censoring levels both Efron's approach and the predicted difference quantity approach $W_{\nu}$ perform similarly to each other in terms of all three indicators. The predicted difference quantity approach performs less well but still better than Efron's approach for $P_{\%}=70$. The MSE of $\beta$ is usually decomposed by bias and variance i.e.~$\mbox{MSE}(\hat{\beta})=\mbox{Var}(\hat{\beta})+[\mbox{Bias}(\hat{\beta})]^2$. If the bias is large, the bias then dominates the MSE. This may explain why the predicted difference quantity approach attains the highest MSE for $\beta$ in two lower censoring levels.

The following simulation example is conducted particularly to understand how the effects of the imputation approaches change over the censoring levels and different correlation structures.
\begin{table}\centering
\caption{Summary statistics for the first simulation example $r_{ij}$ = 0.5. Comparison between the imputation approaches $W_0$: Efron's redistribution, $W_{\tau_{m}}$: conditional mean, $W_{\tau_{md}}$: conditional median, $W_{\tau^{\ast}_{m}}$: resampling based conditional mean, $W_{\tau^{\ast}_{md}}$: resampling based conditional median, and $W_{\nu}$: predicted difference quantity.} \scalebox{0.50}{
\begin{tabular}{@{}lccccccccccccccccccccc@{}}\toprule\toprule
&\multicolumn{6}{c}{$P_{\%}=30$}&&\multicolumn{6}{c}{$P_{\%}=50$}&&\multicolumn{6}{c}{$P_{\%}=70$}\\
\cmidrule{2-7}\cmidrule{9-14}\cmidrule{16-21}&$W_0$&$W_{\tau_{m}}$&$W_{\tau_{md}}$&$W_{\tau^{\ast}_{m}}$&$W_{\tau^{\ast}_{md}}$&$W_{\nu}$
&&$W_0$&$W_{\tau_{m}}$&$W_{\tau_{md}}$&$W_{\tau^{\ast}_{m}}$&$W_{\tau^{\ast}_{md}}$&$W_{\nu}$&&$W_0$&$W_{\tau_{m}}$&$W_{\tau_{md}}$&$W_{\tau^{\ast}_{m}}$&$W_{\tau^{\ast}_{md}}$&$W_{\nu}$\\
\hline
Bias&&&&&&&&&\\
$\beta_1$&-0.295&-0.306&-0.301&-0.307&-0.303&-0.291&&-0.329&-0.362&-0.345&-0.362& -0.349&-0.329&&-0.387&-0.552&-0.542&-0.561&-0.524&-0.441\\
$\beta_2$&-0.042&-0.071&-0.065&-0.077&-0.070&-0.017&&0.113&0.041&0.060&0.016& 0.038&0.126&&0.251&0.290&0.252&0.104&0.253&0.466\\
$\beta_3$&0.372&0.318&0.334&0.306&0.319&0.290&&0.413&0.364&0.411&0.345&0.376& 0.481&&0.580&0.248&0.311&0.145&0.280&0.533\\
$\beta_4$&0.606&0.555&0.568&0.540&0.552&0.527&&0.652&0.557&0.608&0.539&0.563& 0.689&&1.235&0.814&0.953&0.743&0.905&1.253\\
$\beta_5$&1.015&0.963&0.977&0.943&0.958&0.920&&1.061&0.944&0.997&0.903&0.945& 1.093&&2.256&1.772&1.924&1.809&1.862&1.950\\
\hline
Variance&&&&&&&&&\\
$\beta_1$&0.186&0.184&0.182&0.181&0.181&0.191&&0.254&0.260&0.259&0.268&0.264& 0.243&&1.539&0.896&0.840&0.836&0.787&1.123\\
$\beta_2$&0.165&0.165&0.163&0.164&0.164&0.165&&0.286&0.266&0.247&0.261&0.238& 0.278&&0.382&0.301&0.266&0.536&0.273&0.422\\
$\beta_3$&0.169&0.168&0.166&0.166&0.166&0.185&&0.259&0.245&0.245&0.260&0.252& 0.291&&0.551&0.396&0.357&0.403&0.370&0.610\\
$\beta_4$&0.155&0.157&0.151&0.157&0.152&0.168&&0.312&0.293&0.248&0.258&0.240& 0.338&&0.909&0.519&0.384&0.493&0.391&0.743\\
$\beta_5$&0.169&0.170&0.167&0.171&0.167&0.173&&0.290&0.264&0.241&0.239&0.237& 0.319&&0.714&0.579&0.267&0.556&0.397&0.938\\
\hline
MSE&&&&&&&&&\\
$\beta_1$&0.272&0.276&0.270&0.274&0.271&0.273&&0.360&0.388&0.375&0.396&0.384& 0.350&&1.535&1.111&1.050&1.068&0.983&1.183\\
$\beta_2$&0.165&0.168&0.166&0.168&0.167&0.164&&0.296&0.265&0.248&0.259&0.237& 0.291&&0.407&0.355&0.303&0.493&0.310&0.380\\
$\beta_3$&0.306&0.267&0.276&0.258&0.266&0.353&&0.492&0.375&0.412&0.377&0.390& 0.519&&0.832&0.418&0.418&0.384&0.411&0.659\\
$\beta_4$&0.520&0.463&0.472&0.447&0.455&0.590&&0.795&0.601&0.615&0.546&0.554& 0.810&&2.343&1.130&1.253&0.995&1.171&1.683\\
$\beta_5$&1.198&1.096&1.120&1.058&1.082&1.297&&1.473&1.152&1.232&1.052&1.127& 1.511&&5.732&3.662&3.942&3.773&3.826&4.486\\[1ex]
\bottomrule\bottomrule
\end{tabular}}
\label{tab:weightingEX2}
\end{table}

%

\begin{figure}[h]
\centering
\includegraphics [scale=0.60] {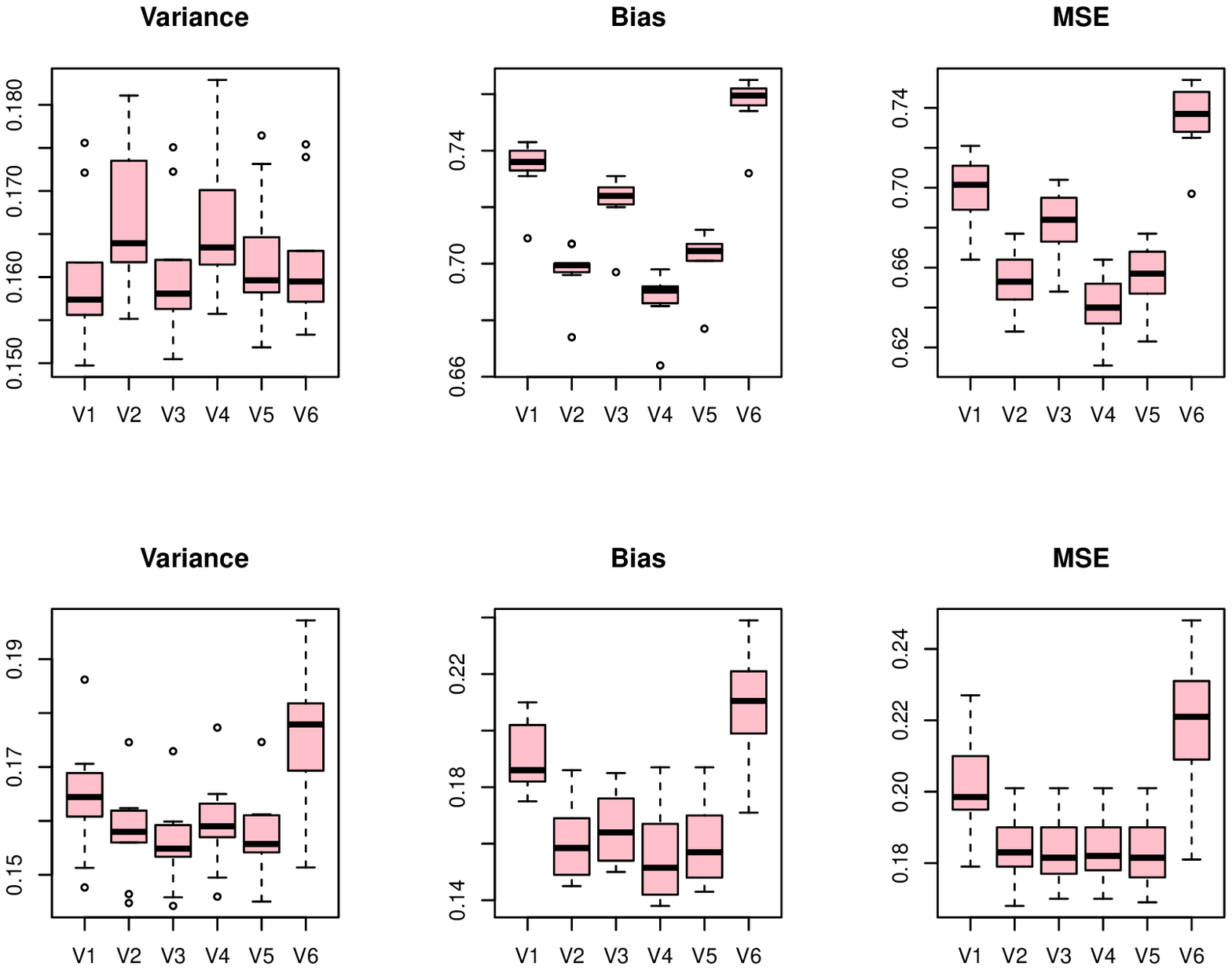}
\caption{Simulation study for second example under $P_{\%}=30\%$.
Box-plots for variance, bias, mean squared error of $\beta$. Here
V1, V2, V3, V4, V5 and V6 represent the estimation approaches $W_0$:
Efron's redistribution, $W_{\tau_{m}}$: conditional mean,
$W_{\tau_{md}}$: conditional median, $W_{\tau^{\ast}_{m}}$:
resampling based conditional mean, $W_{\tau^{\ast}_{md}}$:
resampling based conditional median, and $W_{\nu}$: predicted difference quantity
respectively. Graphs in first row show the results when
$r_{ij}$ is 0; graphs in second row show the results
when $r_{ij}$ is 0.5.} \label{fig:boxweighting30}
\end{figure}

\begin{figure}[h]
\centering
\includegraphics [scale=0.60] {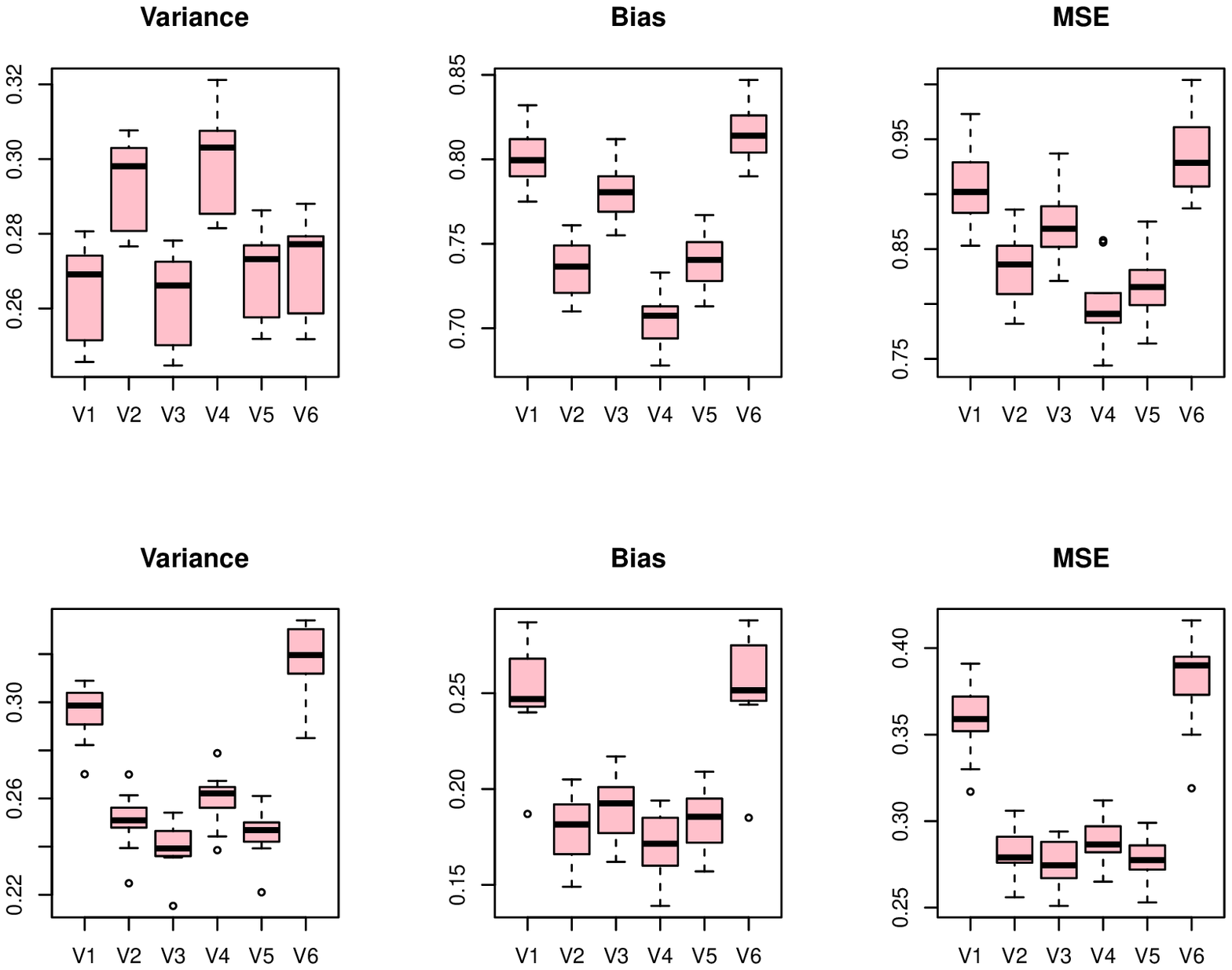}
\caption{Simulation study for second example under $P_{\%}=50\%$.
Box-plots for variance, bias, mean squared error of $\beta$. Here
V1, V2, V3, V4, V5 and V6 represent the estimation approaches $W_0$:
Efron's redistribution, $W_{\tau_{m}}$: conditional mean,
$W_{\tau_{md}}$: conditional median, $W_{\tau^{\ast}_{m}}$:
resampling based conditional mean, $W_{\tau^{\ast}_{md}}$:
resampling based conditional median, and $W_{\nu}$: predicted difference quantity
respectively. Graphs in first row show the results when
$r_{ij}$ is 0; graphs in second row show the results
when $r_{ij}$ is 0.5.} \label{fig:boxweighting50}
\end{figure}

\begin{figure}[h]
\centering
\includegraphics [scale=0.60] {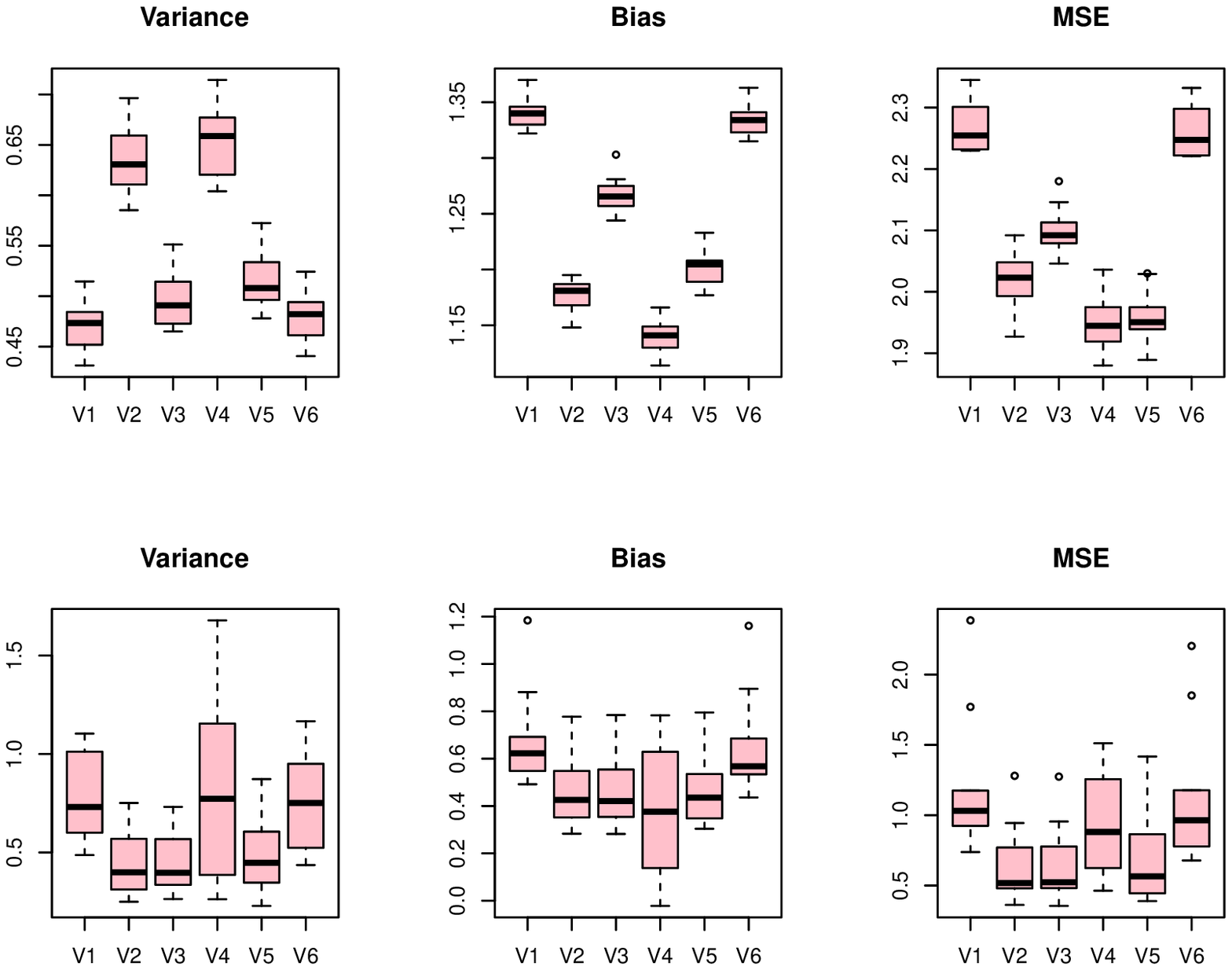}
\caption{Simulation study for second example under $P_{\%}=70\%$.
Box-plots for variance, bias, mean squared error of $\beta$. Here
V1, V2, V3, V4, V5 and V6 represent the estimation approaches $W_0$:
Efron's redistribution, $W_{\tau_{m}}$: conditional mean,
$W_{\tau_{md}}$: conditional median, $W_{\tau^{\ast}_{m}}$:
resampling based conditional mean, $W_{\tau^{\ast}_{md}}$:
resampling based conditional median, and $W_{\nu}$: predicted difference quantity
respectively. Graphs in first row show the results when
$r_{ij}$ is 0; graphs in second row show the results
when $r_{ij}$ is 0.5.} \label{fig:boxweighting70}
\end{figure}

\subsection{Second Example}
We keep everything similar to the previous example except that
$p=10$ and $\beta_j=3$ for $j=1,\cdots,p$.
The results are presented as summary box plots.
Figures \ref{fig:boxweighting30} to \ref{fig:boxweighting70}
represent the results corresponding to three censoring levels $P_{\%}$:
$30$, $50$, and $70$.
The results for this example are similar
to the results of the first example.
The results for the uncorrelated datasets as shown in Figure \ref{fig:boxweighting50}
suggest that adding the resampling based conditional mean
gives the lowest bias and the lowest MSE, but yields the highest variance.
For correlated datasets the other four approaches give lower variance, bias
and MSE than both the Efron's approach and the predicted difference quantity approach.
It is also noticed from the correlated data analysis that the median based approaches
i.e.~adding the conditional median or adding the resampling based conditional median
perform slightly better than the mean based approaches
i.e.~adding the conditional mean or adding the resampling based conditional mean.

\section{Real Data Examples}
We present two well-known real data examples.
The analysis for the first example is done with the larynx cancer data and for the second example, the analysis is done using the Channing House data. The Channing House data is different from the larynx data since the data has heavy censoring and also has many largest censored observations.

\subsection{Larynx Data}
This example uses hospital data where 90 male patients were diagnosed with cancer of the larynx, treated in the period 1970--1978 at a Dutch hospital [Kardaun (1983\nocite{Kard:83:Statis})].
An appropriate lower bound either on the survival time (in years) or on the censored time (whether the patient was still alive at the end of the study) was recorded.
Other covariates such as patient's age at the time of diagnosis,
the year of diagnosis, and the stage of the patient's cancer were also recorded.
Stage of cancer is a factor that has four levels,
ordered from least serious (I) to most serious (IV).
Both the stage of the
\begin{table}[ht]\centering
\caption{Parameter estimation under the approaches $W_0$: Efron's redistribution, $W_{\tau_{m}}$: conditional mean, $W_{\tau_{md}}$: conditional median, $W_{\tau^{\ast}_{m}}$: resampling based conditional mean, $W_{\tau^{\ast}_{md}}$: resampling based conditional median, and $W_{\nu}$: predicted difference quantity approach to the Laryngeal cancer data. Estimates under LN-AFT are based on log-normal AFT model solved by least squares method without tail
correction (see Klein and Moeschberger (1997).} \scalebox{0.6}{
\begin{tabular}{@{}lccccccc@{}}\toprule\toprule
&\multicolumn{7}{c}{Parameter Estimate}\\
\cmidrule{2-8}Variable&$W_0$&$W_{\tau_{m}}$&$W_{\tau_{md}}$&$W_{\tau^{\ast}_{m}}$&$W_{\tau^{\ast}_{md}}$&$W_{\nu}$&LN-AFT\\
\hline
$X_1: \mbox{Age}$&0.008 (0.020)&0.009 (0.022)&0.009 (0.022)&0.009 (0.024)&0.009 (0.021)&0.008 (0.019)&-0.018 (0.014)\\
$X_2: \mbox{Stage II}$&-0.628 (0.420)&-0.846 (0.539)&-0.840 (0.514)&-1.052 (0.535)&-0.966 (0.500)&-0.649 (0.468)&-0.199 (0.442)\\
$X_3: \mbox{Stage III}$&-0.945 (0.381)&-1.176 (0.443)&-1.169 (0.419)&-1.395$^{*}$ (0.451)&-1.304$^{*}$ (0.458)&-0.967 (0.390)&-0.900 (0.363)\\
$X_4: \mbox{Stage IV}$&-1.627$^{**}$ (0.461)&-1.848$^{**}$ (0.444)&-1.841$^{**}$ (0.495)&-2.056$^{**}$ (0.506)&-1.969$^{**}$ (0.581)&-1.648$^{**}$ (0.478)&-1.857$^{**}$ (0.443)\\
\bottomrule\bottomrule
\end{tabular}}
\label{tab:weightingLarynxdata}\\\begin{footnotesize}$(\cdot)^{**}$ and $(\cdot)^{*}$ indicate significant at $p<0.001$ and $p<0.01$ respectively\end{footnotesize}
\end{table}
cancer and the age of the patient were a priori selected as important variables possibly influencing the survival function. We have therefore, $n=90$, $p=4$ ($X_1$: patient's age at diagnosis; $X_2$: 1 if stage II cancer, 0 otherwise; $X_3$: 1 if stage III cancer, 0 otherwise; $X_4$: 1 if stage IV cancer, 0 otherwise). The censoring percentage $P_{\%}$ is 44 and the  largest observation is censored (i.e.~$Y_{(n)}^+=10.7+$).
The dataset is also analysed using various approaches such as log-normal AFT modeling in Klein and Moeschberger (1997\nocite{Klei:Moes:97:book}).

We apply the proposed imputation approaches to the log-normal AFT model (\ref{eq:exampleW})
using regularized WSL (\ref{eq:wlseW3}).
We use two main effects, age and stage.
The estimates of the parameters under different imputation techniques
are reported in Table \ref{tab:weightingLarynxdata}.
These give broadly similar results, but differ from those
found by Klein and Moeschberger (1997\nocite{Klei:Moes:03:book})
where Efron's tail correction was not applied,
shown in the last column of the table (LN-AFT).
Klein and Moeschberger (1997\nocite{Klei:Moes:03:book}) found Stage IV
to be the only significant factor influencing the survival times.
All our imputation methods found Stage IV as highly significant factor. In addition, Stage III factor is found as significant at $p<0.01$ by adding the resampling based mean and median methods.


\subsection{Channing House Data}
Channing House is a retirement centre in Palo Alto, California. The data were collected between the opening of the house in 1964 and July 1, 1975. In that time 97 men and 365 women passed through the centre. For each of these, their age on entry and also on leaving or death was recorded. A large number of the observations were censored mainly due to the residents being alive on July 1, 1975, the end of the study.
It is clear that only subjects with entry age smaller than or equal to age on leaving or death can become part of the sample. Over the time of the study 130 women and 46 men died at Channing House. Differences between the survival of the sexes was one of the primary concerns of that study.

Of the 97 male lifetimes,
51 observations were censored and the remaining 46 were observed exactly.
Of the 51 censored lifetimes, there are 19 observations each of which
has lifetime 137 months (which is the largest observed lifetime).
Similarly, of 365 female lifetimes,
235 observations were censored and the remaining 130 were observed exactly.
Of the 235 censored lifetimes, 106 take the maximum observed value of 137 months.
Therefore, the imputation approaches impute the lifetime of 19 observations
for the male dataset and 106 observations for the female dataset.

\begin{figure}[h]
\centering
\includegraphics [scale=0.50] {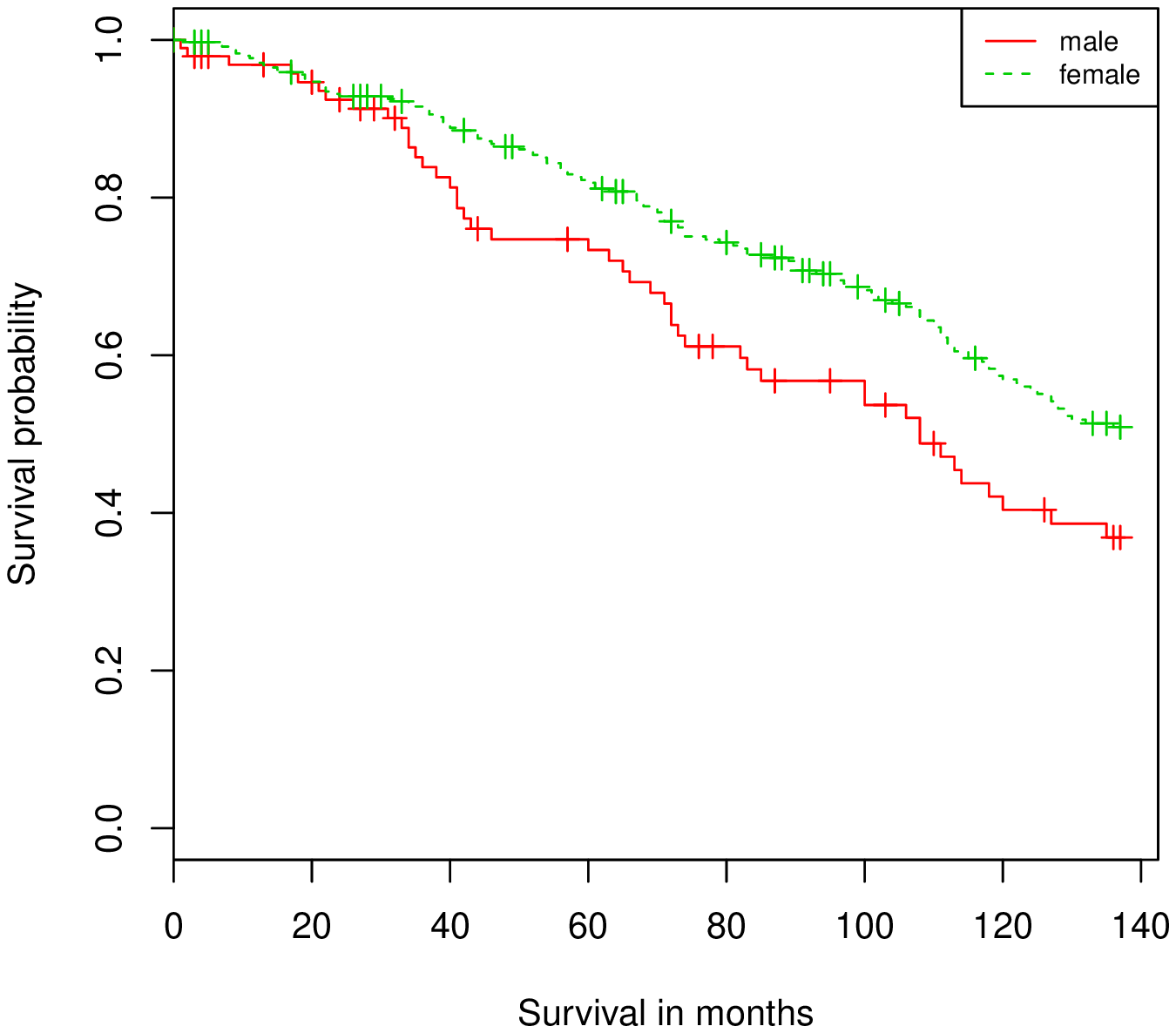}
\caption{K--M plots for Channing House data} \label{fig:kmplot.CannData}
\end{figure}

The K--M survival curve for male and female data,
(Figure \ref{fig:kmplot.CannData})
shows that survival chances clearly differ between the sexes.
We now investigate whether the imputed value and the estimate from the log-normal AFT model (\ref{eq:exampleW}) of lifetimes on the calender ages (the only covariate)
fitted by the WLS method (\ref{eq:wlseW}) differ between male and female.
Of interest we implement the imputing approaches except the resampling based conditional mean and conditional median approaches for male and female Channing House data separately.
The two resampling based approaches can not be implemented for AFT models with one single covariate.
They need at least two covariates to be executed.
The results are shown in Table \ref{tab:ChanningHouseData}.
The results clearly depict that the estimates for age by the methods differ significantly between male and female.
So does happen also for the imputed values obtained by the methods. For both datasets the conditional mean approach imputes with much higher value.

\begin{table}[h]\centering
\caption{Parameter estimation for calender age from the log-normal AFT model and imputed value estimation for the largest observations by the approaches $W_0$: Efron's redistribution, $W_{\tau_{m}}$: conditional mean, $W_{\tau_{md}}$: conditional median, and $W_{\nu}$: predicted difference quantity approach for the Channing House data.} \scalebox{1}{
\begin{tabular}{@{}lcccc@{}}\toprule\toprule
&$W_0$&$W_{\tau_{m}}$&$W_{\tau_{md}}$&$W_{\nu}$\\
\hline
$\mbox{Age (male)}$&-0.153& -0.201& -0.154& -0.154\\
$\mbox{Age (female)}$&-0.180& -0.198& -0.182& -0.186\\
$\mbox{Imputed value (male)}$&137*&176.5&138.1&137.9\\
$\mbox{Imputed value (female)}$&137*&143.1&137.6&138.8\\
\bottomrule\bottomrule
\end{tabular}}
\label{tab:ChanningHouseData}\\
Note: The value with * is not imputed rather than\\ obtained using Efron's redistribution algorithm.
\end{table}


Here we also note that all imputed methods impute the last largest censored observations with a single value. Hence, in the present of heavy censoring the proposed two additional approaches--\emph{Iterative} and \emph{Extrapolation} as described in Section \ref{subs:addTail} can easily be implemented. Both techniques are implemented to male and female Channing House data separately. We report here results only for male data (Table \ref{tab:ChanningHouseAdiImp}).

%
%

\begin{table}[ht]\centering
\caption{Results by the additional imputation methods for the Channing House male data.} \scalebox{0.8}{
\begin{tabular}{@{}ll@{}}\toprule\toprule
Method&Imputed lifetimes for 19 tail tied observations\\
\hline
Iterative method&137.85, 138.11, 138.19, 138.22, 138.22, 138.23, 138.23, 138.23, 138.23, 138.23,\\& 138.23, 138.23, 138.23, 138.23, 138.23, 138.23, 138.23, 138.23, 138.23\\
\hline
Extrapolation method& 134.23, 136.23, 150.25, 152.25, 156.26, 158.26, 160.26, 162.27, 166.27, 176.29,\\&
178.29, 180.29, 184.30, 186.30, 188.30, 194.31, 196.32, 198.32, 200.32\\
\bottomrule\bottomrule
\end{tabular}}
\label{tab:ChanningHouseAdiImp}
\end{table}

We apply the extrapolation based additional imputing method for tied largest censored observations as stated in Section \ref{sssection:extrapolation} to the Channing House data, where in first attempt we find a linear trend between the K--M survival probabilities $\hat{S}(t)$ and the lifetimes. As part of the remaining procedure we first fit a linear regression of lifetimes on $\hat{S}(t)$ and then compute the predicted lifetime against each censored lifetime. The imputed values obtained by using the extrapolation method are put in ascending order in Table \ref{tab:ChanningHouseAdiImp}. Results show that the extrapolation method tends to impute the largest observations with a huge variation among the imputed values. The method also doesn't impute any tied values. The method produces the imputed values in a way as if the largest censored (also tied) observations have been observed. On the contrary, the iterative method imputes values with many ties. In this case all imputed values are close to the imputed value 137.9 that is obtained by the predicted difference approach.

\begin{figure}[ht]
   \includegraphics[scale=0.4]{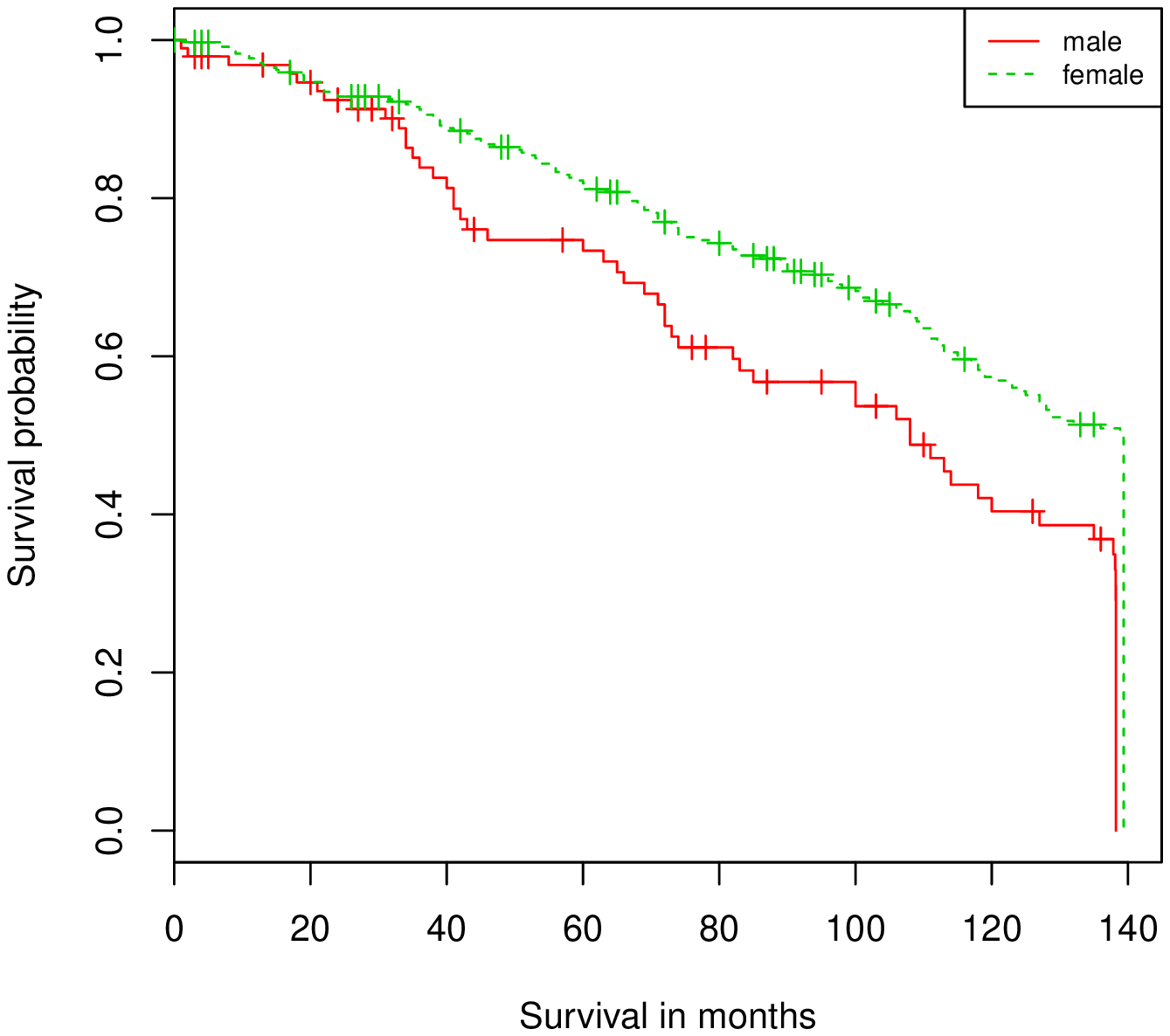}
   \includegraphics[scale=0.4]{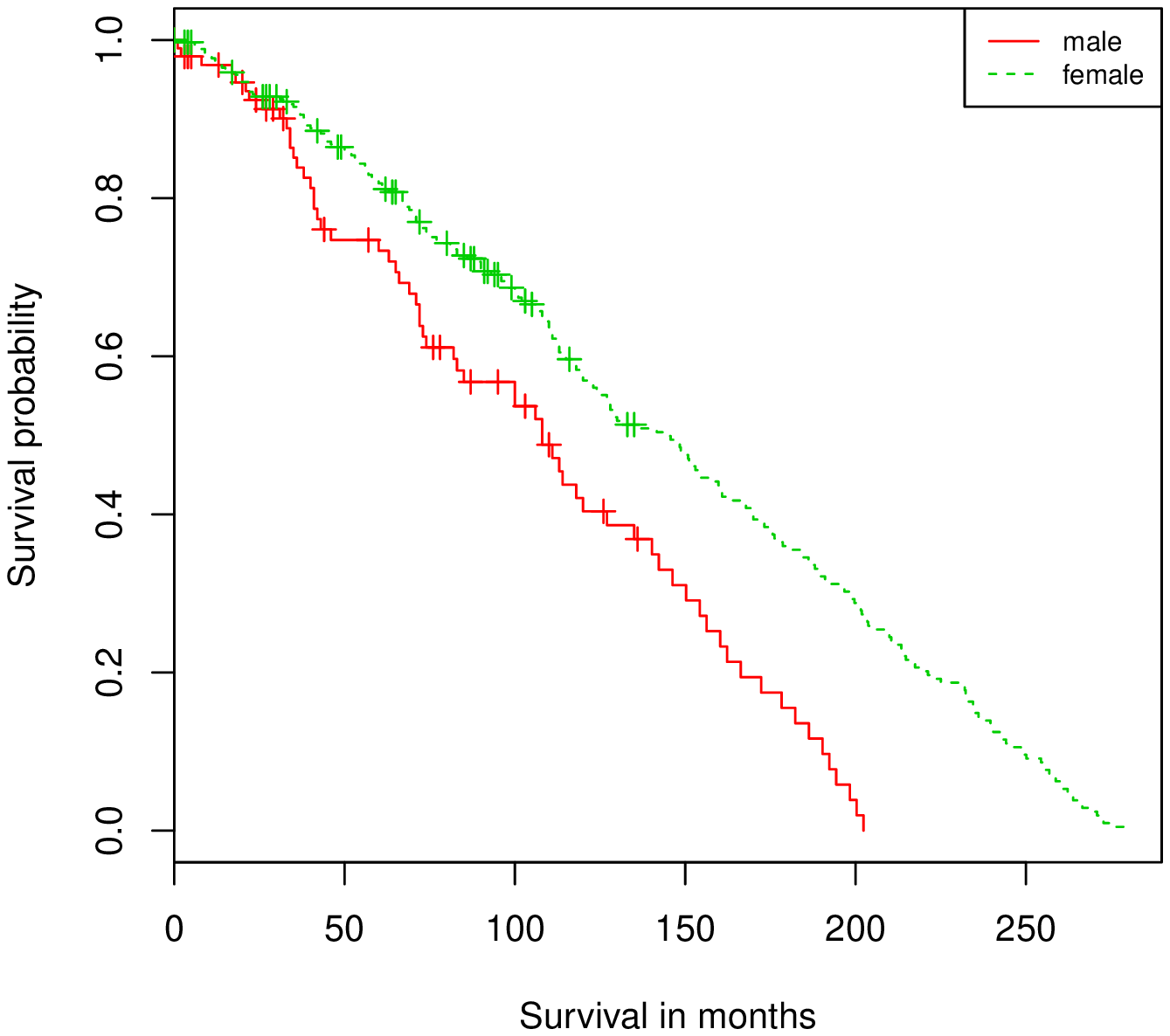}
\caption{Survival comparison between the male and female for Channing House data using the iterative (left panel) and extrapolation (lower panel) additional imputing methods to the tail tied observations}
\label{fig:kmplot2.CannData}
\end{figure}
The K--M plot with the 19 imputed lifetimes for male and 106 imputed lifetimes for female data under two imputing approaches is given by Figure \ref{fig:kmplot2.CannData}. The K--M plots show that the second method outperforms the first by a huge margin. This also leads to major changes in the coefficient value for the age covariate from fitting the AFT model. The estimated coefficient for male data using the two methods are $-0.154$ and $-0.218$ and those for female data are $-0.187$ and $-0.385$. Hence it might suggest that when there are many largest censored observations the second method prefers to the first for imputing them under the AFT models fitted by the WLS method. The first method might be useful when there are very few largest censored observations.

\section{Discussion}
We propose five imputation techniques for the largest
censored observations of a dataset.
Each technique satisfies the basic right censoring assumption
that the unobserved lifetime is greater than the observed censored time.
We examine the performance of these approaches by taking into account different
censoring levels and different correlation structures among the
covariates under log-normal accelerated failure time models.
The simulation analysis suggests that all five imputation techniques except the predicted difference quantity can perform much
better than Efron's redistribution technique for both type of
datasets---correlated and uncorrelated. At higher censoring the predicted difference quantity approach outperforms the Efron's technique while at both lower and medium censoring they perform almost similar to each other.
For both type of datasets,
the conditional mean adding and the resampling based
conditional mean adding provide the least bias and the least mean
squared errors for the estimates in each censoring level.
In addition to the five approaches, we also propose two additional imputation approaches to impute the tail tied observations. These approaches are investigated with two real data examples. For implementing all proposed imputation approaches we have provided a publicly available package \texttt{imputeYn} (Khan and Shaw, 2014\nocite{has:ewa:Rpack:imputeYn}) implemented in the R programming system.
%
%


\section{Acknowledgements}
The first author is grateful to the centre for research in Statistical Methodology (CRiSM), Department of Statistics, University of Warwick, UK for offering research funding for his PhD study.

\newpage
\bibliographystyle{spbasic}      
\bibliography{b2ndyear}   


%
%

\end{document}